\documentclass[aps, prl, showpacs, superscriptaddress, twocolumn]{revtex4}
\usepackage{graphicx}
\usepackage{hyperref}

\begin{document}

\title{Excited States at Interfaces of a Metal-Supported Ultrathin Oxide Film}

\author{T. Jaouen}
\altaffiliation{Corresponding author.\\ thomas.jaouen@unifr.ch}
\affiliation{D{\'e}partement de Physique and Fribourg Center for Nanomaterials, Universit\'e de Fribourg, CH-1700 Fribourg, Switzerland}

\author{E. Razzoli}
\affiliation{D{\'e}partement de Physique and Fribourg Center for Nanomaterials, Universit\'e de Fribourg, CH-1700 Fribourg, Switzerland}

\author{C. Didiot}
\affiliation{D{\'e}partement de Physique and Fribourg Center for Nanomaterials, Universit\'e de Fribourg, CH-1700 Fribourg, Switzerland}

\author{G. Monney}
\affiliation{D{\'e}partement de Physique and Fribourg Center for Nanomaterials, Universit\'e de Fribourg, CH-1700 Fribourg, Switzerland}

\author{B. Hildebrand}
\affiliation{D{\'e}partement de Physique and Fribourg Center for Nanomaterials, Universit\'e de Fribourg, CH-1700 Fribourg, Switzerland}

\author{F. Vanini}
\affiliation{D{\'e}partement de Physique and Fribourg Center for Nanomaterials, Universit\'e de Fribourg, CH-1700 Fribourg, Switzerland}

\author{M. Muntwiler}
\affiliation{Swiss Light Source, Paul Scherrer Institute, CH-5232 Villigen PSI, Switzerland}

\author{P. Aebi}
\affiliation{D{\'e}partement de Physique and Fribourg Center for Nanomaterials, Universit\'e de Fribourg, CH-1700 Fribourg, Switzerland}

\begin{abstract}

We report layer-resolved measurements of the \textit{unoccupied} electronic structure of ultrathin MgO films grown on Ag(001). The metal-induced gap states at the metal/oxide interface, the oxide band gap as well as a surface core exciton involving an image-potential state of the vacuum are revealed through resonant Auger spectroscopy of the Mg $KL_{23}L_{23}$ Auger transition. Our results demonstrate how to obtain new insights on \textit{empty} states at interfaces of metal-supported ultrathin oxide films.

\end{abstract}

\pacs{79.60.Dp, 68.47.Gh, 73.20.-r, 78.70.Dm}
\date{\today}
\maketitle

Metal-supported ultrathin oxide films are a class of materials of technological importance in various research fields such as catalysis, spintronics, or nanoelectronics \cite{Freund2008, Netzer2010}. Their unique chemical and physical properties have raised questions on the role played by reduced dimensionality and the nature of interactions at the  metal/oxide interface \cite{Freysoldt2007}. In this context, MgO/Ag(001) is a model system of the metal/oxide interface at the ultrathin limit. Although the structure and the growth mechanism \cite{Pal2014, Jaouen2014, Luches2004, Valeri2002, Schintke2001, Wollschlager1998}, as well as changes in electronic properties associated with depositing ultrathin films of MgO on Ag(001) have been investigated \cite{Droubay2014, Jaouen2012, Jaouen2010, Bieletzki2010, Konig2009, Prada2008, Giordano2006, Butti2004}, capturing the physical nature of such a mixed system, and in particular of the interfaces, remains challenging. 

In Resonant Auger Spectroscopy (RAS), the Auger process can be very different from that occurring with photon energies far above absorption thresholds \cite{Brown1980}. Sub-lifetime narrowing effects \cite{Liu1994}, as well as strong modulations in Auger signals \cite{Sparks1974}, can occur. Furthermore, depending on whether or not, the resonantly excited electron delocalizes to the conduction band before the core-hole decay, the decay process can result in a two-hole ($2h$) final state ("normal" Auger decay) or in a two-hole and one electron ($2h1e$) final state (spectator channel of the autoionization process), respectively. These two competing decay pathways are both visible in resonant Auger spectra if the time scale of delocalization is comparable to the core-hole lifetime. Thus, information on the screening of the core hole, the degree of localization of excited electrons, or the charge transfer dynamics at interfaces and surfaces can be obtained \cite{Keller1998, Bruhwiler2002, Fohlisch2005, Menzel2008}.

In this letter, we study the evolution of the layer-resolved Mg $KL_{23}L_{23}$ Auger transition for a 3 monolayer (ML) thick MgO film grown on Ag(001), in a photon energy range corresponding to the Mg $K$-edge. In good agreement with density functional theory (DFT) calculations, we show that the intensity evolution of the resonant Auger spectra with the photon energy allows to get a layer-by-layer mapping of the local density of empty Mg $p$-states probed by the excited photoelectron in the intermediate dipole transition \cite{Drube1995}. We find that, in the pre-edge region, the Auger spectra mostly consist of a \textit{single} Auger component, the one of the metal/oxide interface, demonstrating the metallic character of the oxide interface layer due to the presence of metal-induced gap states (MIGS). We measure the MgO surface band gap and a spectroscopic fingerprint of a surface core exciton involving an image potential state of the vacuum.
 
All experiments were performed at the Photoemission and Atomic Resolution Laboratory (PEARL) beam line situated at bending magnet X03DA of the Swiss Light Source. The MgO films were grown \textit{in-situ} on Ag(001) (for details see Ref. \onlinecite{Jaouen2014}). X-ray absorption was measured by recording the total electron yield (TEY) and RAS was obtained at room temperature using a VG Scienta EW4000 with 60$^{\circ}$ acceptance angle for photon energy steps of 0.2 eV in the range of 1300-1320 eV. The total energy resolution was 0.7 eV. The DFT calculations have been performed within the full potential linearized augmented plane wave formalism implemented in the WIEN2K package \cite{WIEN2K} using the modified Becke-Johnson (mBJ) exchange-correlation potential \cite{Tran2009, footnote}. The MgO/Ag (001) system was modeled by a 15 layer Ag(001) surface slab covered on both sides by 3 ML of MgO with lattice parameter $a_0=$ 4.16 \AA~ and Ag interface atoms below the oxygen anions. The vacuum region between adjacent slabs was set to $\sim$ 40 \AA~. 
   
\begin{figure}[t]
\includegraphics[width=0.45\textwidth]{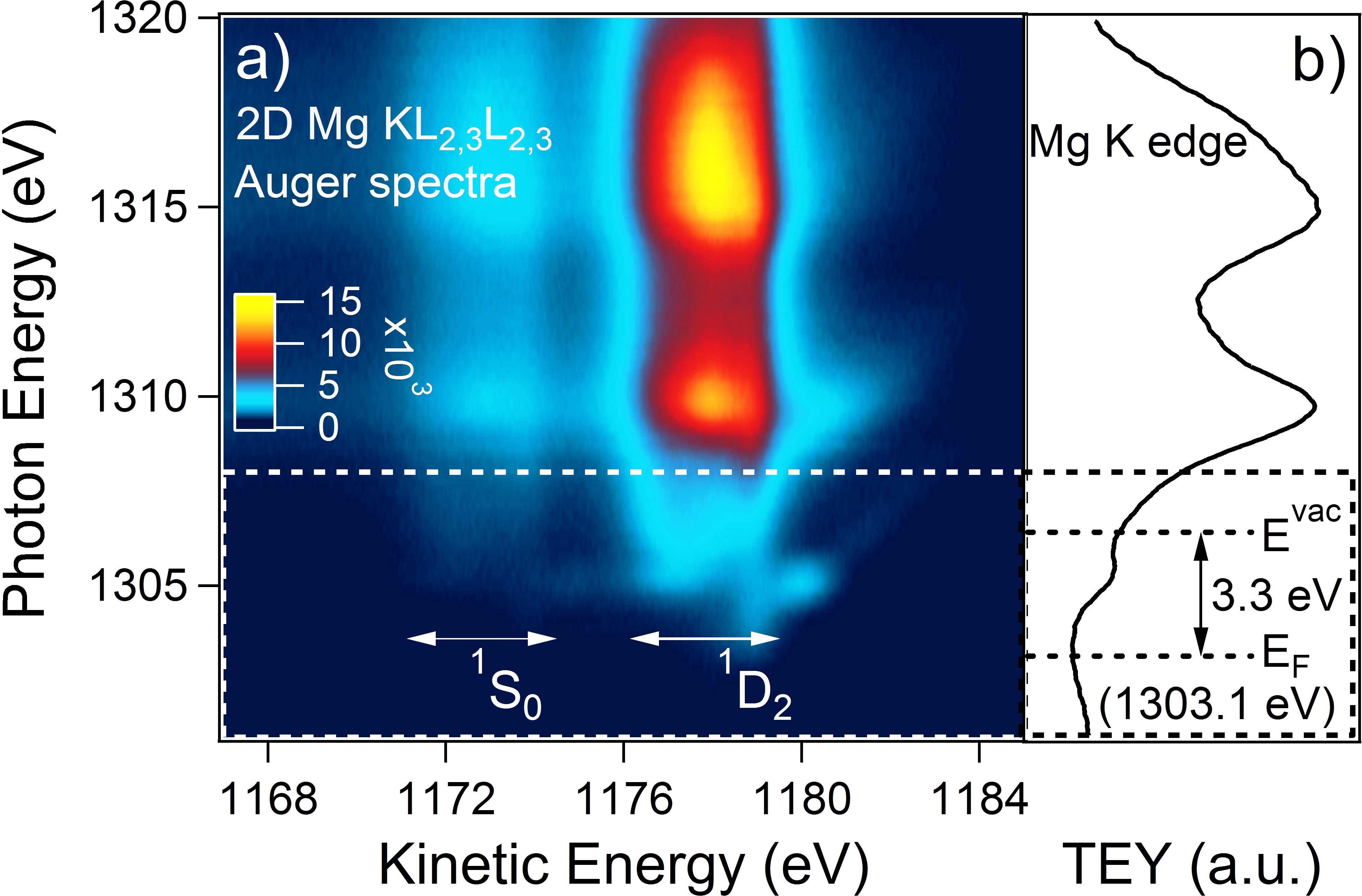}
\caption{\label{fig1} (color online). (a) Two-dimensional (2D) color-scaled intensity map of the Mg $KL_{23}L_{23}$ Auger transition for photon energies ranging from 1301 eV to 1320 eV, corresponding to the Mg $K$-absorption edge. (b) Total electron yield (TEY) while the excitation energy is scanned through the Mg $K$-edge absorption threshold. The positions of the Ag Fermi edge ($E_F$) and vacuum level ($E^{vac}$) are indicated.}
\end{figure}

Figure \ref{fig1}(a) and \ref{fig1}(b) respectively show the Mg $KL_{23}L_{23}$ Auger transition intensity obtained by scanning the photon energy across the Mg $1s$ $\rightarrow$ $3p$ x-ray absorption resonance and the corresponding TEY spectrum for a MgO(3ML)/Ag(001) sample. Excitation to the Fermi level ($E_F$) occurs at a photon energy of 1303.1 $\pm$ 0.2 eV [see Fig. \ref{fig1}(b)]. This value is the Mg $1s$ binding energy obtained by x-ray photoelectron spectroscopy, measuring the energy difference between the silver Fermi edge and the Mg $1s$ core level. The first absorption maximum which lies at 1310.4 $\pm$ 0.1 eV photon energy, represents an excitation to states 7.3 $\pm$ 0.3 eV above $E_F$. Knowing that $E_F$ lies 3.85 $\pm$ 0.10 eV above the MgO valence band maximum (VBM) and that the MgO/Ag(001) work function value (defined as the energy difference between the vacuum level $E^{vac}$ and $E_F$) is 3.30 $\pm$ 0.05 eV  \cite{Jaouen2010}, the first strong resonance maximum at 1310.4 eV corresponds to electron excitation into the continuum. In this continuum region, the Auger spectra consisting of the 5 eV-separated $^1S$ and $^1D$ multiplet of the Mg $2p$ final state, show intensity enhancement at the two strongest resonance maxima. Although slightly shifted and distorted by post-collision interaction \cite{Armen1987}, they have a constant kinetic energy as expected for a \textit{normal} Auger decay and for excitations above the ionization threshold. 

\begin{figure}[b]
\includegraphics[width=0.45\textwidth]{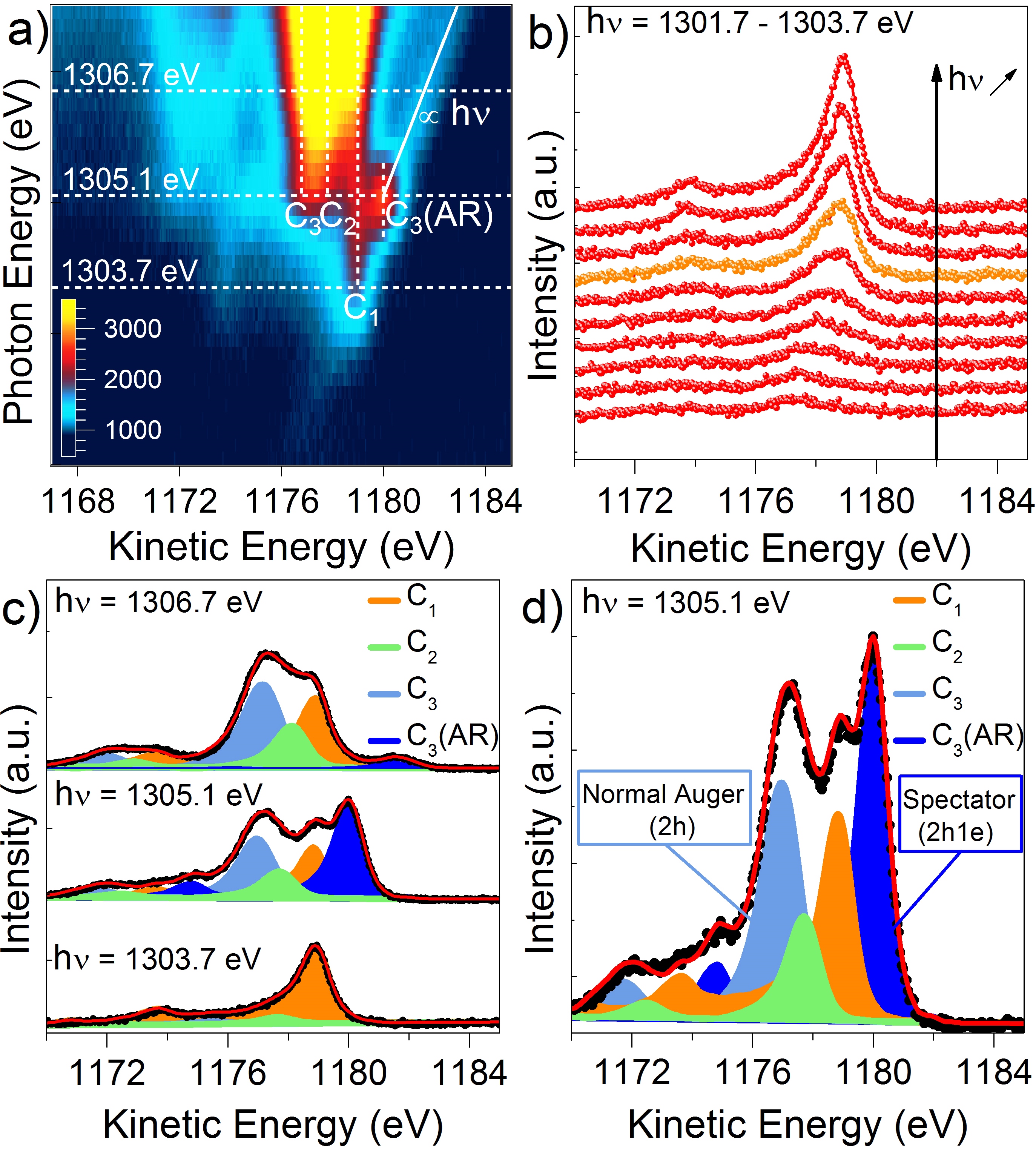}
\caption{\label{fig2} (color online).(a) Detailed intensity map of the Mg $KL_{23}L_{23}$ Auger transition between 1301 eV and 1308 eV photon energy. The white line corresponds to a constant binding energy line. (b) Waterfall plot of the resonant Auger spectra in the 1301-1303.7 eV region. The orange curve indicates the Auger spectrum obtained for $h\nu$= 1303.1 eV, i.e. excitation to $E_F$. (c) Resonant Auger spectra corresponding to the three horizontal line profiles at 1303.7, 1305.1, and 1306.7 eV (white-dashed lines) in (a). Best fit and layer-by-layer decomposition are also shown. (d) Resonant Auger spectrum obtained with 1305.1 eV photon energy.}
\end{figure}

More interesting is the Mg $K$ pre-edge [region enclosed by the white-dashed rectangle in Fig. \ref{fig1}(a)], with the excitation of the $1s$ electron into the MgO band gap region. Figure \ref{fig2}(a) shows an intensity map of the Mg $KL_{23}L_{23}$ Auger transition between 1301 eV and 1308 eV photon energy. The vertical dashed-white lines (labeled $C_1$, $C_2$, $C_3$) correspond to the energy positions of the $^1D_2$ multiplet associated with the interface ($C_1$), sub-surface ($C_2$), and surface ($C_3$) Auger emission of the layer-resolved Mg $KL_{23}L_{23}$ Auger transition \cite{Jaouen2013}. Starting at 1301.7 [Fig. \ref{fig2}(b)], i.e. 1.4 eV below $E_{F}$, we observe a linear dispersion of the kinetic energy of the Auger transition with photon energy and a gradual distortion of its lineshape together with a quick increase of the intensity. Around $E_{F}$ [orange curve in Fig. \ref{fig2}(b)], the normal Auger transition develops with the $^1D_2$ multiplet centered at 1179.0 eV. This peculiar behavior is characteristic of Auger resonant Raman scattering in metals, describing the transition from the resonant Auger-Raman (AR) regime to the normal Auger (NA) one at $E_{F}$ \cite{Drube1995}. One remarkable result to be pointed out at this stage is that these Auger spectra essentially consist of a \textit{single} Auger component, the interface one ($C_1$) [Fig. \ref{fig2}(c), see bottom spectrum]. In this photon energy region, the only available $p$-unoccupied states for the Mg 1$s$ electron are those of the interface layer. 

With the increase of the photon energy, $C_2$ and $C_3$ appear [Fig. \ref{fig2}(a), horizontal line at 1305.1 eV], as well as an additional surface contribution \cite{footnote1}, on the high-energy side of the main Auger peaks. The intensity of this surface contribution [$C_3$(AR)], reaches its maximum at 1305.1 eV photon energy [Fig. \ref{fig2}(c), center spectrum]. It shows a resonant Auger-Raman behavior [see the constant binding energy line on Fig. \ref{fig2}(a)] and is the spectator channel of the autoionization process which leads to a $2h1e$ final state. Its kinetic energy is larger than the one associated with the normal surface Auger transition ($C_3$) by $\sim$3 eV due to the additional screening interaction of the core hole with the excited electron [Fig. \ref{fig2}(d)].  

\begin{figure}[t]
\includegraphics[width=0.45\textwidth]{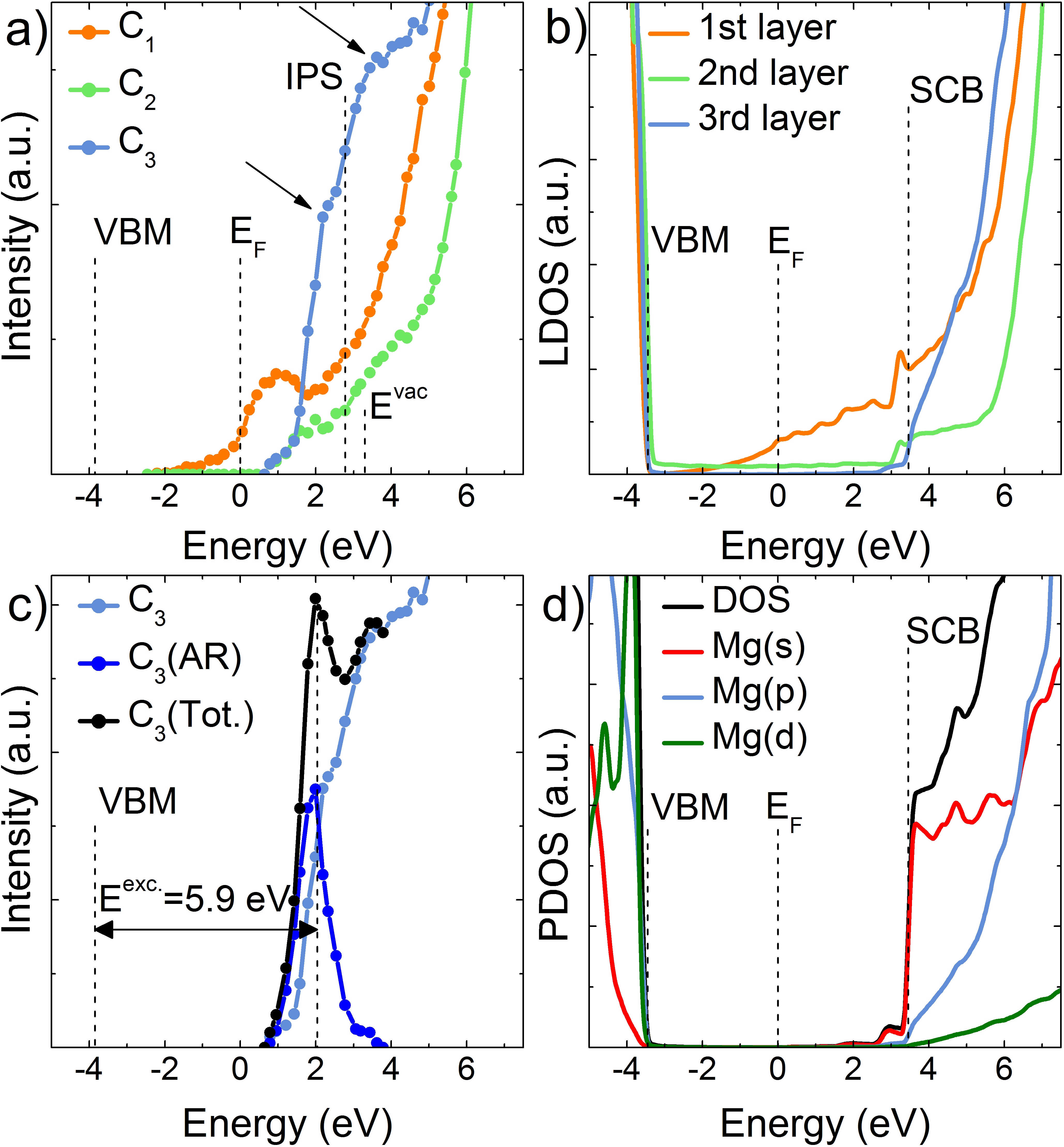}
\caption{\label{fig3} (color online). (a) Intensity evolution of $C_1$, $C_2$ and $C_3$ as a function of the energy relative to E$_F$. The positions of $E^{vac}$, VBM and of the image potential state (IPS) of the MgO surface are indicated (see text). The arrows shows the maxima of the two rising edges of $C_3$. (b) Local density of Mg $p$-states (LDOS) as calculated by DFT-mBJ. The calculated surface conduction band (SCB) edge is also indicated. (c) Intensity evolution of the surface contribution [$C_3$(Tot.)] decomposed into normal Auger ($C_3$) and Auger-Raman [$C_3$(AR)] decay channels. (d) Partial density of states (PDOS) of Mg surface atoms.}
\end{figure}

To quantify these experimental findings, we performed curve fitting analysis as depicted in Fig. \ref{fig2}(c) and (d). The fitting procedure used the experimental Mg $KL_{23}L_{23}$ Auger spectrum obtained at $h\nu$= 1303.5 eV constituted of the single interface component \cite{footnote2}. Fig. \ref{fig3}(a) shows the intensity evolution of $C_1$, $C_2$ and $C_3$ as a function of the energy relative to $E_F$. The agreement with DFT-mBJ-calculated local density of Mg $p$-states (LDOS) [Fig. \ref{fig3}(b)] is satisfactory thus demonstrating that the intensity of the resonant Auger spectra is, for \textit{each} MgO layer, modulated by the unoccupied DOS probed in the intermediate dipole transition \cite{Drube1995}. In particular, in nice agreement with the calculated interface LDOS [orange curve in Fig. \ref{fig3}(b)], the $C_1$ Auger signal [Fig. \ref{fig3}(a)] is non-zero at $E_F$ and throughout the band gap region, directly demonstrating that the MgO interface layer is \textit{metallic}. Furthermore, the MgO band gap already develops for the second layer [green curve, Fig. \ref{fig3}(a)]. The surface curve [blue curve, Fig. \ref{fig3}(a)], which exhibits a "two-step-like" shape (see arrows) and which is more intense than the ones of the interface and sub-surface layers, is not fully reproduced by DFT-mBJ. 

Figure \ref{fig3}(c) focuses on the intensity evolution of the surface contributions. The total intensity associated with the surface [$C_3$(Tot.)] reveals a well resolved sharp Lorentzian-shaped peak 2.05 $\pm$ 0.10 eV above $E_F$, very similar to what is experimentally obtained for the C $1s$ absorption edge in diamond and is a typical spectroscopic fingerprint of a core exciton \cite{Morar1985}. Therefore, between $\sim$1 and 2 eV above $E_F$, the first rising edge of the $C_3$ evolution [up to the first arrow in Fig. \ref{fig3}(a)] and the $C_3$(AR) Auger emission are the NA and the AR channels of a surface core exciton decay, respectively. The intensity associated with $C_3$(AR) is centered at 5.9 $\pm$ 0.2 eV with respect to the MgO VBM and corresponds to the surface core exciton energy [Fig. \ref{fig3}(c)]. The $C_3$ contribution which is resonantly enhanced together with $C_3$(AR) is of fully excitonic origin, thus explaining that it can not be described within DFT, and that the first rising edge of $C_3$ in Fig. \ref{fig3}(a) is missing in Fig. \ref{fig3}(b). 

Next, we can see that the surface LDOS shows a continuous increase starting with a small step 3.49 eV above $E_F$ [blue curve in Fig. \ref{fig3}(b)]. As seen in Fig. \ref{fig3}(d), at this energy the total DOS (black curve) shows a clear step characteristic of a 2D-DOS associated with the presence of an unoccupied surface state of mainly $s$-character (red curve). Although hardly visible when projected onto the $p$-orbitals, this step is clearly present in experiment and corresponds to the \textit{second} rising edge of the $C_3$ evolution of Fig. \ref{fig3}(a) (between the two arrows). This indicates that the DFT-mBJ is not well suited for capturing the physical nature of this unoccupied surface state.     

\begin{figure}[t]
\includegraphics[width=0.45\textwidth]{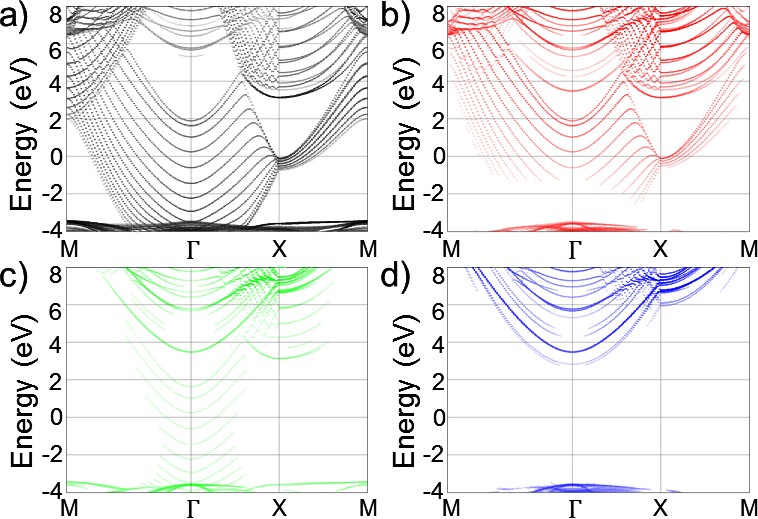}
\caption{\label{fig4} (color online). DFT-calculated band structures of the MgO(3ML)/Ag(001) slab projected onto the interface silver orbitals (a), and on the interface (b), sub-surface (c), and surface (d) Mg $p$-orbitals. The energy reference is taken at the Fermi level.}
\end{figure}

For a deeper understanding of this discrepancy, of the non-zero DOS in the band gap region of the interface layer, and of the electronic origin of the conduction band edge and of the surface core exciton, we calculate the layer-resolved band structure in DFT. In Fig.\ref{fig4} the band structure of the MgO(3ML)/Ag(001) system is projected onto the interface Ag orbitals, and on the interface, sub-surface, and surface Mg $p$-orbitals. The Ag interface layer clearly shows the localized 4$d$ energy bands lying $\sim$4 eV below $E_F$ and the nearly free-electron like 5$sp$ bands which are partly occupied [Fig. \ref{fig4}(a)]. Looking at the oxide side of the interface, we see that the Ag 5$sp$ states hybridize with the Mg 3$p$ orbitals as in-gap states [Fig. \ref{fig4}(b)]. They penetrate into the oxide film and are quickly damped going towards the MgO surface [Fig. \ref{fig4}(b)-(d)]. 

This behavior is in good agreement with the experimental results of Fig. \ref{fig3}(a) and is characteristic of MIGS \cite{Heine1965, Louie1976}. These states are the tails of the metal wave functions that decay exponentially into the insulator and take primarily their weight from the bands that are nearest in energy \cite{Tersoff1984, Bordier1991}. For MgO, they are constituted of the O 2$p$ states close to the valence band and of the Mg 3$s$-$p$ states close to the conduction band, as seen in Fig. \ref{fig4}(b). We thus conclude that the pure interface signal obtained in the Mg $K$ pre-edge region, results from the Auger decay following the excitation of Mg 1$s$ electrons into MIGS. Compared to previous experiments \cite{Muller1998, Schintke2001, Kiguchi2003}, the power of RAS resides in the ability of selectively probing MIGS even if they are localized at the metal/oxide buried interface. 

Focusing on the MgO surface band structure, we see that the DFT-mBJ surface band gap of 6.94 eV is the energy distance between the top of the O $2p$ bands (3.45 eV below $E_F$, results not shown) and an unoccupied surface state 3.49 eV above $E_F$ at $\Gamma$ [Fig. \ref{fig4} (d)]. It has been theoretically shown that this DFT-calculated surface state was indeed an image potential state (IPS) located in the vacuum, outside the surface, when described by the most accurate many-body perturbation theory \cite{Rohlfing2003}. Compared to the surface state which is localized on the surface cations and which mainly derives from Mg 3$s$ states, the IPS is delocalized in the surface plane and exhibits free-electron dispersion parallel to the surface. Whereas the excitation of an 1$s$ electron into a surface state is constrained by the dipole selection rule, the excitation into an IPS is not forbidden. This explains that the second rising edge of the $C_3$ evolution in Fig. \ref{fig3} (a) exhibits a clear 2D-DOS shape which is not reproduced by the DFT-calculated surface LDOS, and clearly indicates that the experimental surface conduction band (SCB) edge is an IPS. Considering the maximum slope of the second rising edge of $C_3$ in Fig. \ref{fig3}(a), the IPS lies 0.52 $\pm$ 0.15 eV below $E^{vac}$ and its energy relative to the VBM is 6.63 $\pm$ 0.2 eV, values in nice agreement with those obtained in $GW$ calculations \cite{Rohlfing2003}. 

Finally, we note that our measured surface core exciton lies 0.7 $\pm$ 0.2 eV below the IPS. Furthermore, from the extracted Auger and autoionization intensities of the surface components [Fig. \ref{fig3}(c)] , we can get the ratio $C_{3}$(AR)/$C_{3}$. By applying the core-hole-clock (CHC) method \cite{Menzel2008}, which uses the Mg $1s$ core hole lifetime of 1.88 fs as an internal clock \cite{Citrin1977}, we finally obtain the surface core-exciton lifetime. It reaches $\sim$5 fs at the resonance maximum and is 2.7 times larger than the Mg 1$s$ core-hole lifetime. Considering the surface core exciton energy with respect to the IPS, together with the rather short lifetime, this is strong evidence that the excited electron involves IPS and that it is localized in all three dimensions at the site of the core hole. While pure IPS are delocalized in the surface plane, the electron of the surface core exciton which is bound, both by the image potential and by Coulomb interaction with the hole, is also trapped laterally. For MgO, this hybrid of an IPS and an exciton has been previously predicted by Cox \textit{et al.} \cite{Cox1986}, and calculated at a similar binding energy \cite{Rohlfing2003}. It has been also observed at the surface of an organic semiconductor by two-photon photoemission spectroscopy \cite{Muntwiler2008}, and is expected for any insulating or semiconducting surface where the hole is incompletely screened.     

To summarize, through the selective probing of the MIGS, surface band gap and core exciton of the MgO(3ML)/Ag(001) system, we have obtained a complete view of the unoccupied electronic structure of a metal-supported ultrahin oxide film. Whereas information on such electronic states are hardly obtained in conventional photoemission-based techniques, our RAS study further demonstrates how to capture the electronic properties of a single layer embedded in a thicker film. We believe that the conclusions drawn in this paper are not restricted to the MgO/Ag(001) system and could be extended to other metal-oxide combinations, thus opening new possibilities for determining their \textit{whole} electronic structure with thickness sensitivity.   

\begin{acknowledgments}
This project was supported by the Fonds National Suisse pour la Recherche Scientifique through Div. II. The authors warmly acknowledge B. Hediger, F. Bourqui and O. Raetzo for technical support.  
\end{acknowledgments}


\begin{thebibliography}{1}
\bibitem{footnote2} The experimental spectra were reproduced by keeping fixed the kinetic energies of $C_1$, $C_2$, and $C_3$. The only varied parameters between the different spectra were the kinetic energy of $C_3$(AR) and the intensities of all components resulting in a very robust fit.
\end{thebibliography}
\end{document}